
\font\twelvebf=cmbx12
\font\ninerm=cmr9
\nopagenumbers
\magnification =\magstep 1
\overfullrule=0pt
\baselineskip=18pt
\line{\hfil IASSNS-HEP-94/88}
\line{\hfil CCNY-HEP 94/9}
\line{\hfil October 1994}
\vskip .3in
\centerline{\twelvebf Many-body States and Operator Algebra }
\centerline{\twelvebf for Exclusion Statistics}
\vskip .4in
\centerline{\ninerm DIMITRA KARABALI}
\centerline{Institute for Advanced Study}
\centerline{Princeton, NJ 08540}
\vskip .4in
\centerline{\ninerm V.P. NAIR}
\centerline{ Physics Department}
\centerline{City College of the City University of New York}
\centerline{New York, New York 10031}
\vskip 1in
\baselineskip=16pt
\centerline{\bf Abstract}
\vskip .1in
We discuss
many-body states and the algebra of creation and annihilation operators
for particles obeying exclusion statistics.
\vskip .5in
\footnote{}{E-mail addresses: karabali@sns.ias.edu,
vpn@ajanta.sci.ccny.cuny.edu}
\vfill\eject
\baselineskip=20pt
\parskip=0pt
\def\dag{\dagger}
\def\a{\alpha}

\def\d{\delta}     \def\D{\Delta}

\def\l{\lambda}
\def\o{\omega}
\def\s{\sigma}
\def\bra{\langle}
\def\ke{\rangle}
\def\af2{\vert f_2 \vert}
\footline={\hss\tenrm\folio\hss}
\pageno=1
Recently Haldane introduced a variant of fractional statistics for which the
notion of exclusion, a generalization of the Pauli principle, rather than
exchange or braiding properties, is the prime characteristic$^{[1]}$. This
exclusion
statistics appears naturally in the fractional
quantum Hall effect, spin-${1 \over 2}$ antiferromagnetic chains and
the Calogero-Sutherland model$^{[1-5]}$.

Exclusion principle implies that the number of available one-particle states
should change with increasing occupation of a state and hence exclusion
statistics can be characterized by the change $\D d_i$ in available states as
the occupation number is changed by $\D N_i$, i.e., by
$$ \D d_i = - \sum_{j} g_{ij} ~ \D N_j \eqno (1) $$
$g_{ij}$ are the parameters characterizing the statistics. $d_i$ is the number
of one-particle states available to the $N_i$-th particle with quantum numbers
$i$, holding the labels of the ($N_i -1$) particles fixed. Particles obeying
exclusion statistics were called $g$-ons in ref.[6], a name we shall also use.

The thermodynamic distributions for particles
obeying exclusion statistics can be set up generalizing the long familiar
combinatorial calculation of entropy for bosons and fermions. A number of
thermodynamic properties have been studied in this framework$^{[4-6]}$. A
question
which naturally arises in this context is whether one can
go beyond the thermodynamic formulation, attempting a microscopic
description by explicitly
constructing the many-body Hilbert space and further
introducing creation
and annihilation operators and appropriate commutation rules.

In this paper we discuss an algebra of creation and annihilation
operators which realizes some of the general features following from
Eq.(1). The operator algebra is different
from algebras discussed in connection with the Calogero-Sutherland
system$^{[7]}$; further it
seems to be different from proposals for anyon commutation rules$^{[8]}$ as
well as $q$-deformations of boson and fermion algebras, although there are many
similarities$^{[9,10]}$.

We start by considering $g$-ons of a single energy $\o$, i.e., only one
species.
Let $|n \ke $ denote an $n$-body state of $g$-ons.
(We consider them to be in a
cavity of volume $V=L^3$ and
fixed momentum of the form ${{2 \pi} \over L} \times {\rm integer}$).
$n$ can be taken as the
eigenvalue of some hermitian operator $N$. Further, we can assume $\bra n|m
\ke = \d _{n,m}$.

We now introduce an annihilation operator $a$ by
$$ a |n \ke = f_n |n-1 \ke \eqno (2) $$
By orthonormality of states, $\bra n-1|a|n \ke = f_n $, which leads, by the
definition of the adjoint, to
$$ a^{\dag} |n \ke = f^{*}_{n+1} |n+1 \ke \eqno (3)  $$
For a sensible annihilation operator we need $a |0 \ke =0 $, or $f_0 =0$ and
the one-particle state can be defined as $a^{\dag} | 0 \ke $, so that $f_1 ^{*}
=1$.

The number operator $N$, by definition, must obey the commutation rules
$$ [N,a]=-a, ~~~~~~~~~[N, a^{\dag}]=a^{\dag} \eqno(4) $$
from which it follows that $[N, a^{\dag}a]= [N, a a^{\dag}]=0$.
The number operator $N$ can thus be written, in general, as
a function of $a^{\dag}a$ or $a a^{\dag}$.

{}From Eq.(1) we see that
for $g=1/m$, $m$ being an integer,
we lose a state every time the occupation number increases
by $m$. We thus consider that in a microscopic description, Eq.(1)
implies that the maximal occupancy of the state is $m$.
In terms of operators this is equivalent to the condition
$(a^{\dag})^{m+1}=(a)^{m+1}=0$. A basis for the Hilbert space consists of the
states $\vert 0 \ke,~ \vert 1 \ke,..., \vert m \ke $.
By definition the number operator $N$ satisfies the polynomial relation
$N(N-1)(N-2)...(N-m)=0$. Further it
can be written as an $m$-th order polynomial in $a^{\dag} a$ (or $a a^{\dag}$),
$$ N = \sum _{k=1} ^{m} \l _k (a^{\dag} a)^k \eqno (5) $$
Applying this on $|0 \ke , |1 \ke , |2 \ke$, etc., and using
Eqs.(2) and (3) we get
the equations
$$ \sum_{k=1} ^{m} |f_n|^{2k} \l _k = n , ~~~~~~~~~~~~n=1,2,...,m
\eqno (6) $$
We can now solve these equations for $\l _k$ and hence
obtain $N$ as a function of $\vert f_l \vert ^2, ~l=2,...,m$.
Similarly we can write down a ``commutation rule" expressing the relation
between $a a^{\dag}$ and $a^{\dag} a$,
$$ a a^{\dag} = 1 + \sum_{k=1}^{m} \s _k (a^{\dag} a )^k \eqno (7) $$
where $\s _k$'s satisfy the equations
$$ \sum_{k=1} ^{m} \vert f_n \vert ^{2k} \s _k = \vert f_{n+1} \vert ^2 -1
\eqno
(8) $$
The general solution to Eqs.(6) and (8) is given by
$$ \eqalignno{
\l _k & = \sum _{l=1} ^{m} {l \over k!} {1 \over {detM}}
\left[ {{\partial ^k} \over {(\partial x_l)
^k}} detM \right]_{x_l =0}  \cr
\s _k & = \sum _{l=1} ^{m} {{(x_{l+1}-1)} \over k!} {1 \over {detM}}
\left[ {{\partial ^k} \over {(\partial x_l)
^k}} detM \right]_{x_l =0}  &(9) \cr}
$$
where
$$ detM = -(x_1 x_2 ...x_m) \prod_{i<j} (x_i - x_j) \eqno(10) $$
and $M_{lk} = (x_l)^k$. We substitute $x_l = |f_l|^2 $
after evaluating the right hand side of Eq.(9).

For example, for $m=2$ (``semions") we find
$$ \eqalignno{
a a^{\dag} & = 1+ {{\af2 ^6 - \af2 ^4 +1} \over {\af2 ^4 -\af2 ^2}} ~a^{\dag} a
-
{{\af2 ^4 - \af2 ^2 +1} \over {\af2 ^4 - \af2 ^2}} ~(a^{\dag} a)^2 \cr
& = 1 - (1-\vert f_2 \vert ^2) ~a^{\dag} a -
{{1+ ( \vert f_2 \vert ^2-1) \vert f_2 \vert ^2 } \over {\vert f_2 \vert ^2}}
{}~(a^{\dag})^2 a^2 &(11) \cr
N & = {{\af2 ^4 -2} \over {\af2 ^4 - \af2 ^2}} ~a^{\dag} a + {{2 - \af2 ^2}
\over
{\af2 ^4 - \af2 ^2}} ~(a^{\dag}a)^2 \cr
& = a^{\dag} a + {{2 - \vert f_2 \vert ^2} \over { \vert f_2 \vert ^2}}
{}~(a^{\dag})^2 a^2 &(12) \cr}
$$
We further have $(a^{\dag})^3 =0 ,~ a^3 = 0$.
(The
case $m=2$ is similar to the case considered in ref.[9].)

Eqs.(5-10) give an ($m$-1)-parameter set of operator algebras, the
parameters being $x_l=\vert f_l \vert ^2,~~l=2,...,m$.
Specializations of this class of algebras can be obtained by imposing
relations among $\vert f_l \vert ^2$. Such relations can be written in
the form of a recursion rule
$$ \vert f_{n+1} \vert ^2 = K(n,g) \vert f_n \vert ^2 , ~~~~n=1,...
\eqno(13) $$
the choice of the function $K$ now parametrizing the possible algebras.
Once a $K(n,g)$ is chosen, Eq.(13) can give, upto a phase, a recursive
determination of $f_n$'s and hence a representation for
$a$ and $a^{\dag}$. Some restrictions on $K(n,g)$ follow from general
arguments. In order to obtain $f_{m+1}=0$ from Eq.(13), consistent with
the maximum-$m$ occupancy for a state, $K(n,g)$ must vanish for $n=m=1/g$.
Thus we may expect a factor like $(1-gn)$ to some positive power. Further
we have the limiting cases
$K(n,g)  = {{1-n} \over n}$ for $g=1$ and
$K(n,g) = {{1+n} \over n}$ for $g=0$,
corresponding to fermions and bosons respectively.
$\vert f_{n+1} \vert ^2$ is essentially the probability of introducing an
extra $g$-on into a state with occupation number $n$. For $m$ very large and
$n/m$ small, the system is very nearly bosonic and
the exclusion effect is insignificant; we expect
$\vert f_{n+1} \vert ^2 \sim (1-gn) + n $, corresponding to a spontaneous
emission term proportional to the available states $(1-gn)$ and a
stimulated emission term $\sim n$.
Examples of possible choices for $K(n,g)$ are
$$ \eqalignno {
K(n,g) & = {{(1-gn) [1+(1-g)n]} \over n} &(14a) \cr
K(n,g) & = {{(1-gn)^g [1+(1-g)n]^{1-g}} \over n} &(14b) \cr}
$$
There are of course many other choices possible.
We expect $\vert f_{n+1} \vert ^2 / \vert f_n \vert ^2 $ to be
proportional to the number of
ways of introducing an extra $g$-on into the $n$-filled state, which is
related to the combinatorial entropy per particle, say $S(n)$, as $\sim
e^{S(n)}$. The combinatorial rule introduced in ref.[4] gives the
entropy as
$$ \eqalignno {
S & = \log \left[ {{1+(1-g)n} \over {1-gn}} \right] + n S(n) \cr
S(n) & = \log \left[ {{(1-gn)^g [1+(1-g)n]^{1-g}} \over n} \right] &(15)
\cr }
$$
Eq.(15) suggests the choice (14b). However we would
like to emphasize that there is no compelling argument for such a choice.
In particular the thermodynamic description of ref. [4] cannot be
derived from a microscopic theory in the standard way, given
that there is no positive density matrix associated with
it$^{[6]}$.
We cannot therefore expect to recover Wu's thermal distribution from
Eq.(14b).
(Note that the maximum occupancy being $m$ is sufficient to ensure that
the thermodynamic
distribution for the number density $\bar{n} \rightarrow m$
as the temperature goes to zero.)
Nevertheless we shall analyze choice (14b), and its appropriate
generalization in the case of many species, in some detail below because
it is consistent with expected limiting behaviours and interpolates
smoothly between bosons and fermions. Other choices will give qualitatively
similar features.

A further restriction on $K(n,g)$ is derived from the fact that
$\vert f \vert ^2$, being probabilities, cannot be negative.
A closer inspection of Eq.(14b) reveals that there are inconsistencies for some
values of the parameter $g$. The inconsistency appears because, for some values
of $g$, one can eventually encounter negative $\vert f \vert ^2$.
Our approach from now on will be that we shall use Eq.(14b) only for
$g= 1/m$.

Using the recursion rule (14b) we find
$$ \eqalignno{
f_1 & = 1 \cr
|f_n|^2 & =  \prod _{k=1} ^{n-1} {{(1-{k \over m})^{{1 \over m}}
(1+k(1-{1 \over m}))^{1-{{1
\over m}}}} \over k}, ~~~~~~~~~2 \le n \le m  &(16a) \cr
& =  0,  ~~~~~~~~~~~~~~~~n \ge m+1 & (16b) \cr}
$$
As we mentioned earlier, this provides, upto a phase,
a matrix realization of $a$ and $a^{\dag}$; $a_{mn}=\d _{m,n-1}~
f_n,~~~a^{\dag} _{mn} =\d _{m,n+1} ~f_{n+1}^{*}$.

We now turn to the case of many species corresponding to different values of
energy or other quantum numbers. For two species of $g$-ons we
introduce creation and annihilation operators $a_i^{\dag}, a_i,~ i=1,2$ and
$f_{n_1,n_2}^{(i)}$ with
$$ \eqalignno{
& a_1 | n_1,n_2 \ke = f^{(1)} _{n_1,n_2} | n_1 -1, n_2 \ke , \cr
& a_2 | n_1,n_2 \ke = f^{(2)} _{n_1,n_2} | n_1 , n_2 -1 \ke , & (17) \cr}
$$
where $f^{(1)}_{0,n_2}=f^{(2)}_{n_1,0}=0$ and $f^{(1)}_{1,0} = f^{(2)}_{
0,1}=1$.
Eq.(3) can be generalized to the case of two-species as
$$ \eqalignno {
|f^{(1)} _{n_1 +1, n_2} |^2 & = K^{(1)} (n_1, n_2, g)
|f^{(1)} _{n_1,n_2}|^2, ~~~~~~ n_1 > 0 \cr
|f^{(2)} _{n_1 , n_2 +1} |^2 & = K^{(2)} (n_1, n_2, g)
|f^{(2)} _{n_1,n_2}|^2, ~~~~~~ n_2 > 0 &(18) \cr} $$
A suitable generalization of Eq.(14b) is
$$ \eqalignno {
& K^{(i)}(n_1, n_2, g) = (1+w_i) \prod _{k} ({w_k \over {1+w_k}})^
{g_{ki}} \cr
& w_i n_i + g_{ij} n_j = 1 &(19) \cr} $$
The many-body states and the representation of $a_i$ and $a^{\dag}_i$ can
be constructed by solving these recursion relations.
We shall first address the case where $g_{ij}$ is diagonal, i.e., no mutual
statistics, and for the parameters $g_{ii} =1/m$ we shall obtain the full
operator algebra. We shall later discuss the case with mutual statistics
and the somewhat complicated structure of the corresponding many-body Hilbert
space.

In the absence of mutual statistics, the recursion rules (18) reduce to
$$ \eqalignno {
|f^{(1)} _{n_1 +1, n_2} |^2 & = w_1 ^{g_1} (1+w_1)^{1- g_1} |f^{(1)}
_{n_1,n_2}|^2 \cr
|f^{(2)} _{n_1 , n_2 +1} |^2 & = w_2 ^{g_2} (1+w_2)^{1- g_2}
|f^{(2)} _{n_1,n_2}|^2
&(20) \cr} $$
where $g_i= g_{ii}$ and $n_i w_i = 1-g_i n_i$.
For consistency reasons outlined earlier in the case of one species of $g$-ons,
we shall consider special values of $g_i$, in particular the case
$g_1=g_2=1/m$.
We again find that the maximal occupancy for each
species is $m$; i.e., $(a_i)^{m+1} = (a_i^{\dag})^{m+1} =0$. Further, the
recursion rules (20) are sufficient to show that the Hilbert
space is a product of the Hilbert spaces for each species. These properties
clearly generalize to arbitrary number of species.

The recursion rules (20) do not completely determine the functions $f^{(i)} _{
n_1,n_2}$. Specifically, $|f^{(1)} _{1, n_2}|$ and $|f^{(2)} _{n_1, 1}|$ for
$n_1,~n_2 \ne 0$, as
well as the phases of all the $f$'s, are undetermined. Thus Eq.(20)
does not suffice to get all the matrix elements of $a_i$ and $a_i ^{\dag}$.
This
is, of course, no different from the situation with bosons or fermions; one
needs commutation rules relating operators for different species in order to
obtain a complete operator algebra and determine
all the matrix elements.

The
rationale for the choice of off-diagonal commutation rules has to come from the
exclusion principle in the following way.
Although not specified explicitly, we have been dealing all along with free
$g$-ons. Thus the quantum number labelling the species is the spatial momentum
$p$ with a corresponding energy $\o (p)$. The property of exclusion reads, for
$g_i =
1/m $, $(a_p)^{m+1}=(a_p^{\dag})^{m+1}=0$. Eventhough the $p$-diagonal
representation is appropriate for free particles, we can equally well use the
coordinate representation. Since the coordinates $x$ also provide a complete
set of quantum numbers for the particles, we expect the exclusion property
$(a(x))^{m+1}
=0,~ (a^{\dag}(x))^{m+1} =0$ to hold, where $a(x)=\sum_p e^{ipx} a_p$ and
$a^{\dag} (x)$
is the adjoint of $a(x)$. Of course, one could also use $a(q)= \sum_q u_q (p)
a_p
$ in $q$-diagonal representation, $q$'s being any complete set of quantum
numbers and $u_q (p)$ being the appropriate functions. Thus in general we
expect
linear combinations of $a$'s to obey the exclusion principle, i.e.,
$$ (\sum_i \a _i a_i )^{m+1} = (\sum_i \a _i ^{*} a_i ^{\dag} )^{m+1} = 0 \eqno
(21) $$
where $\a _i $ are arbitrary.

{}From the definition of the number operator $N_i \vert n_1,n_2,... \ke = n_i
\vert n_1,n_2,... \ke $, we find that
$$ [N_i, a_j]=[N_i, a_j ^{\dag}]=0, ~~~~~~~~~~i \ne j \eqno(22) $$
Eqs.(21,22) will be our main guide for obtaining the full operator
algebra
for the case of many species. However they are still not sufficient to
determine the algebra completely, so we shall further
assume that
$$ a_i a_j = R_{ij} a_j a_i , ~~~~~~~~~~~~i \ne j \eqno(23) $$
where $R_{ij}$ is a $c$-number. Consistency of Eq.(23) requires $R_{ij} R_{ji}
=1$
. Further, $a_i ^{\dag} a_j ^{\dag} = R^{*} _{ji} a^{\dag} _j a^{\dag} _i$.
Since
$N_i$ is a function of $a_i ^{\dag} a_i$, we seek to satisfy Eq.(22) with the
slightly stronger requirement $[a_i ^{\dag} a_i , a_j]=[a_i ^{\dag} a_i ,
a^{\dag}_j]=0$, $i \ne j$. The general solution to this equation,
using Eq.(23), is
$$ \eqalignno {
a_i ^{\dag} a_j & = R^{-1} _{ij} a_j a^{\dag} _i \cr
a_i a^{\dag} _j & = R^{*}_{ij} a_j ^{\dag} a_i , ~~~~~~~~i \ne j &(24) \cr}
$$
Consistency of these equations gives $R_{ij}^{*} R_{ij} =1$. Thus $R_{ij}$ is a
phase, $R_{ij} = e^{i \theta _{ij}}$. We must now impose Eq.(21). For two
species, we have
$$ (\a _1 a_1 + \a _2 a_2)^{m+1} = (\a _1 a_1)^{m+1} + (\a _2 a_2)^{m+1} +
\sum_{k=1} ^{m} C(k,m,R^{*}_{12}) ((\a _1 a_1)^{m+1-k} (\a _2 a_2)^{k}
\eqno(25) $$
where
$$ C(k,m,R) = {{(1-R^{m+1}) (1-R^m)...(1- R^{m+2-k})} \over
{(1-R) (1-R^2)...(1-R^k)}}, ~~~1 \le k \le m \eqno(26) $$
In particular $C(1,m,R)={{(1-R^{m+1})} \over (1-R)}$. Since
$(a_1)^{m+1}= (a_2)^{m+1}=0$, Eq.(21) can be satisfied if all the $C(k,m,
R^{*}_{12})$ vanish. This determines $R_{12}$. The generic solution for
arbitrary $m$ is of the form $R_{12}
= \exp (\pm 2 \pi i /(m+1))$.
There can be more solutions for special choices of $m$. All solutions are a
subset of the $(m+1)$-th roots of unity. For example, for
$m=2,~R_{12} = e^{\pm 2 \pi i /3}$; $m=3,~ R_{12} = e^{\pm 2 \pi
i /4}$; $m=4,~ R_{12} = e^{\pm 2 \pi
i /5},~e^{\pm 4 \pi i / 5}$. Since we want to treat all $m$'s on an equal
footing, we are going to use the generic solution $R_{12}
= \exp (\pm 2 \pi i /(m+1))$.
The commutation rule becomes $a_1 a_2 = e^{2 \pi i
/(m+1)} a_2 a_1$ or $a_1 a_2 = e^{-2 \pi i/(m+1)} a_2 a_1$.
Notice that the second
choice is obtained from the first by a relabelling $a_1 \leftrightarrow a_2$.
The basic solution will be $R_{12} = e^{2 \pi i /(m+1)}$, up to such
relabellings.

When we have more than two species, the exclusion condition (21) can be
satisfied as follows. Write
$$ a_1 a_3  = R_{13} a_3 a_1, ~~~~~~~~~~~
a_2 a_3  = R_{23} a_3 a_2 \eqno (27) $$
For the choice $R_{23}=R_{13}$,
$$ A a_3 = R_{13} a_3 A $$
where $A= \a_1 a_1 + \a _2 a_2$. The binary expansion (25) suffices to simplify
$(A+\a_3 a_3)^{m+1} =0$. We get the same solution as before, viz.,
$R_{13}=R_{12
}$; the choice $R_{12}=R_{13}=R_{23}$
obviously generalizes inductively to arbitrary number of species.

Recapitulating, the commutation rules we have so far are, for $g_i =
1/m$,
$$ \eqalignno {
a_i a^{\dag}_i & = 1 + \sum_{k=1} ^{m} \s _k (a^{\dag} _i a_i)^k &(28a) \cr
a_i a_j & = \exp ({{2 \pi i} \over {m+1}}) ~a_j a_i ~~~~~~~~i<j &(28b) \cr
a_i^{\dag} a_j^{\dag} & = \exp ({{2 \pi i} \over {m+1}}) ~a_j^{\dag} a_i^{\dag}
{}~~~~~~~~i<j &(28c) \cr
a_i^{\dag} a_j & = \exp (-{{2 \pi i} \over {m+1}}) ~a_j a_i^{\dag} ~~~~~~~~i<j
&(28d) \cr }
$$
We have arrived at Eqs.(28) by using the simplifying ansatz (23), with
$R_{ij}$ being a $c$-number. In order to write these commutation relations
for different
species we have to introduce an ordering, for example $i < j$. The indices
$i,j$, etc., being momentum labels,
ordering is naturally possible only in one spatial dimension. Since many of the
physical situations where exclusion statistics might be relevant are
effectively
one-dimensional$^{[1-5]}$,
this may not be a drastic limitation. It is possible that more
general structures can be constructed taking $R_{ij}$ to be an operator. Notice
also that once we have chosen an ordering of momenta we cannot relabel $a_1
\leftrightarrow a_2$, etc. The commutation rules with $R_{ij} \leftrightarrow
R^{*}_{ij}$ are thus distinct.

We can now write down $f^{(i)} _{n_1,n_2}$ thus providing an explicit
 matrix
realization for $a_i$ and $a_i^{\dag}$. Consider two species, $i=1,2$.
{}From Eqs.(17,28) we derive two sets of equations that should be satisfied by
$f$'s,
$$ \eqalignno {
f^{(2)}_{n_1,n_2} f^{(1)}_{n_1,n_2-1} & = \exp (2 \pi i / (m+1))
f^{(1)}_{n_1,n_2} f^{(2)}_{n_1-1,n_2} &(29a) \cr
f^{*(2)}_{n_1,n_2} f^{(1)}_{n_1,n_2} & = \exp (-2 \pi i / (m+1))
f^{(1)}_{n_1,n_2-1} f^{*(2)}_{n_1-1,n_2} &(29b) \cr}
$$
Eqs. (20,29) completely determine the moduli $\vert f \vert$; in particular,
$$ \eqalignno {
& \vert f^{(1)}_{n_1,n_2} \vert ^2 = \vert f^{(1)}_{n_1,n_2-1} \vert ^2 = ... =
\vert f^{(1)}_{n_1,0} \vert ^2 = \vert f_{n_1} \vert ^2, ~~~~~~0 \le n_1,n_2
\le m
\cr
& \vert f^{(2)}_{n_1,n_2} \vert ^2 = \vert f^{(2)}_{n_1-1,n_2} \vert ^2 = ... =
\vert f^{(2)}_{0,n_2} \vert ^2 = \vert f_{n_2} \vert ^2, ~~~~~~ 0 \le n_1,n_2
\le
m  &(30) \cr }
$$
where $\vert f_{n_i} \vert ^2$ are defined in Eq.(16) for a single
species.

Eqs. (29) also determine some of the relative phases of the $f$'s, but
they still allow some freedom of choice for the phases; this is related
to the ambiguity in how many-particle states are defined. For example,
$(a_1^{\dag })^{n_1} (a_2 ^{\dag})^{n_2} |0 \ke$ and $(a_2 ^{\dag})^{n_2}
(a_1 ^{\dag})^{n_1} |0 \ke $ have
the same number of particles but different phases. We will make the choice
$$ |n_1, n_2,... \ke = C ~ (a_1 ^{\dag})^{n_1} (a_2 ^{\dag})^{n_2}... |0 \ke
\eqno(31) $$
where $C$ is a normalization factor chosen to be real.
This is ordered in terms of ascending momenta; the $a^{\dag}$'s with smaller
labels appear to the left. With this phase convention, we have, for two species
and $m=2$,
$$ \eqalignno {
& f^{(1)}_{1,0}=f^{(1)}_{1,1}=f^{(1)}_{1,2}=f^{(2)}_{0,1}=1 \cr
& f^{(1)}_{2,0}=f^{(1)}_{2,1}=f^{(1)}_{2,2}=f^{(2)}_{0,2}= \left[ {\sqrt{3}
\over 2} \right] ^{1/2} \cr
& f^{(2)}_{1,1}=e^{2 \pi i /3} ~,~~~~~f^{(2)}_{1,2}=e^{2 \pi i /3} \left[
{\sqrt{3} \over 2} \right] ^{1/2} \cr
& f^{(2)}_{2,1}=e^{4 \pi i /3} ~,~~~~~f^{(2)}_{2,2}=e^{4 \pi i /3} \left[
{\sqrt{3} \over 2} \right] ^{1/2} &(32) \cr}
$$
All other $f$'s are zero. We thus have a complete specification of $a_i,~
a_i^{\dag}$. This can be easily extended to many species. In fact, it is
evident that the commutation rules (28) suffice to evaluate the action of
$a_i$'s and $a_i ^{\dag}$'s on any state of the form (31). The exchange
properties of the many-body wavefunctions are evident from the representation
(31) and the commutation rules (28).

In the case of mutual
statistics, viz., $g_{ki}$ not diagonal in Eq.(19), the structure of the
many-body Hilbert space is somewhat more complicated. The number of many-body
states
can be greater or smaller, depending on the signs and magnitudes of $g_{ki}$,
than what is expected from the tensor product of states for each species. This
change
in the expected number of states has to do with Eqs.(18,19). From
Eq.(19), $w_i$ can become negative as $n_j$ increases. Since $|f^{(i)}|^2$ is
positive, a negative $w_i$ can give an inconsistency, as we have
already discussed earlier, unless some of the
$f$'s are zero. Notice that Eq.(18) leaves $f^{(1)}_{1, n_2} , f^{(2)}_
{n_1, 1}$, for $n_1,~n_2 \ne 0$
undetermined. One can set some of these to zero to obtain a consistent
solution to Eqs.(18,19). The possible states are then no longer what is
expected from the tensor product.

We shall illustrate this by considering two characteristic examples
with two species, where the
number of the states increases and decreases respectively, compared
to the tensor product of the single species states.
\vskip .15in
\noindent
A) $g_{11}=g_{22}={1 \over m},~ g_{12}=g_{21}=-{1 \over m}$

In this case the recursion rules reduce to
$$ \eqalignno {
|f^{(1)} _{n_1 +1, n_2} |^2 & = (1+w_1)^{1- {1 \over m}} \left(
{w_1 \over w_2} \right) ^{1 \over m} (1+w_2)^{1 \over m}
|f^{(1)} _{n_1,n_2}|^2 &(33a) \cr
|f^{(2)} _{n_1 , n_2 +1} |^2 & = (1+w_2)^{1- {1 \over m}} \left(
{w_2 \over w_1} \right) ^{1 \over m} (1+w_1)^{1 \over m}
|f^{(2)} _{n_1,n_2}|^2 &(33b) \cr} $$
where $w_1 n_1 = 1 - {(n_1 -n_2) \over m},~w_2 n_2 = 1 + {(n_1 -n_2) \over m}$.
We see that the ratio $w_1 /w_2$ becomes negative if $\vert n_1 -n_2 \vert >
m$.
A consistent solution requires that $f^{(1)} _{1, m+i} =0$ and $f^{(2)}
_{m+i,1}
=0,~i \ge 1$. This choice allows only nonnegative $w_1 / w_2$. (This also
avoids
the singular points $w_2 =0$ in Eq.(33a) and $w_1=0$ in Eq.(33b)).
The allowed states
are of the form $\vert n_1, n_2 \ke$, where $\vert n_1 - n_2 \vert \le m$.
There
are $(2m+1) (m+1)$ such states, which are $m (m+1)$ more than in the tensor
product of $(m+1)$ states for each species.

\vskip 0.15in

\noindent
B) $g_{11}=g_{22}={1 \over m},
{}~ g_{12}=g_{21}={1 \over m}$.

In this case the recursion rules become
$$ \eqalignno {
|f^{(1)} _{n_1 +1, n_2} |^2 & = (1+w_1)^{1- {1 \over m}} \left(
w_1 w_2 \right) ^{1 \over m} (1+w_2)^{-{1 \over m}}
|f^{(1)} _{n_1,n_2}|^2 &(34a) \cr
|f^{(2)} _{n_1 , n_2 +1} |^2 & = (1+w_2)^{1- {1 \over m}} \left(
w_1 w_2 \right) ^{1 \over m} (1+w_1)^{-{1 \over m}}
|f^{(2)} _{n_1,n_2}|^2 &(34b) \cr} $$
where $w_1 n_1 = w_2 n_2 =1 - {(n_1 +n_2) \over m}$. Here we must require
$f^{(1)}_{1, m+i}=0$ and $f^{(2)}_{m+i,1}=0,~ i \ge 0$, so as to avoid
recursive sequences which lead to the
singular points $w_2 =-1$ for Eq.(34a) and $w_1 = -1$ for Eq.(34b).
The allowed states
now have $(n_1 +n_2) \le m$. There are ${1\over 2}(m+2)(m+1)$ such states,
which are ${1\over 2}m(m+1)$ less than in the tensor product of $(m+1)$ states
for each species.

It is clear from the above discussion that the operator algebra for the case of
mutual statistics must be fairly involved so as to reflect the complicated
structure of the many-body states. This is currently under investigation.

\vskip .3 in

$ {\underline {\rm Acknowledgements}}$: The commutation rules (5,7) can
be mapped
into the usual bosonic algebra$^{[11]}$. We thank A.P. Polychronakos for
discussions and for pointing this out.

This work was supported in part by the DOE grant DE-FG02-90ER40542
and the NSF grant PHY-9322591.

\vskip 0.4in
${\underline {\rm Note ~ added}}$: The combinatorial argument after Eq.(14),
with equal a priori probability and maximal occupancy $m$, gives yet
another choice for $K(n, g=1/m)$ as
$$
n={1 \over {K-1}} - {{m+1} \over {K^{m+1} -1}}
$$
It is
unclear how to generalize this equation to the case of mutual
statistics. The case $m=2$ for the above choice of $K$ is similar
to the case considered in ref.[12].
\vskip 0.5 in
\centerline{\bf References}
\noindent
\vskip 0.2in
\item{[1]} F.D.M. Haldane, {\it Phys. Rev. Lett.} {\bf 67} (1991) 937.
\item{[2]} F.D.M. Haldane and B.S. Shastry, in the {\it Proceedings of the 16th
Taniguchi Symposium}, Kashikojima, Japan, October 1993, eds. A. Okiji and N.
Kawakami, Springer-Verlag, 1994.
\item{[3]} Z.N.C. Ha, {\it Phys. Rev. Lett.} {\bf 73} (1994) 1574; preprint
IASSNS-HEP-94/90, cond-mat/9410101; J. Minahan and A. Polychronakos,
{\it Phys. Rev.} {\bf B50} (1994) 4236; F. Lesage, V. Pasquier and D. Serban,
preprint SPhT-94, hep-th/9405008.
\item{[4]} Y.S. Wu, {\it Phys. Rev. Lett.} {\bf 73} (1994) 922.
\item{[5]} D. Bernard and Y.S. Wu, preprint SPhT-94-043, UU-HEP/94-03,
cond-mat/
9404025; A. Dasnieres de Veigy and S. Ouvry, {\it Phys. Rev. Lett.} {\bf 72}
(1994) 600; M.V.N. Murthy and R. Shankar, {\it Phys. Rev. Lett.} {\bf 72}
(1994)
3629; preprint IMSc-94/24, cond-mat/9404096.
\item{[6]} C. Nayak and F. Wilczek, preprint PUPT 1466, IASSNS 94/25, cond-mat/
9405017, to appear in {\it Phys. Rev. Lett.}
\item{[7]} A. Polychronakos, {\it Nucl. Phys.} {\bf B324} (1989) 597;
D.V. Khveshchenko, Princeton University preprint, cond-mat/9404094.
\item{[8]} G. Semenoff, {\it Phys. Rev. Lett.} {\bf 61} (1988) 517;
T. Matsuyama, {\it Phys. Lett.} {\bf B228} (1989) 99; G. Semenoff and P.
Sodano,
{\it Nucl. Phys.} {\bf B328} (1989) 753; D. Karabali, {\it Int. J. Mod. Phys.}
{\bf A6} (1991) 1369.
\item{[9]} A.Y. Ignat'ev and V.A. Kuz'min, {\it Sov. J. Nucl. Phys.} {\bf 46}
(1987) 786.
\item{[10]} O.W. Greenberg and R.N. Mohapatra, {\it Phys. Rev. Lett.} {\bf 59}
(1987) 2507; {\it Phys. Rev. Lett.} {\bf 62} (1989) 712; O.W. Greenberg, {\it
Phys. Rev. Lett.} {\bf 64} (1990) 705; R.N. Mohapatra, {\it Phys. Lett.} {\bf
B242} (1990) 407.
\item{[11]} A.P. Polychronakos, {\it Mod. Phys. Lett.} {\bf A5} (1990) 2325.
\item{[12]} P. Mitra, preprint SISSA-191/94/EP, hep-th/9411236.

\end